\documentclass[
 reprint,
superscriptaddress,
 amsmath,amssymb,  
 aps,
]{revtex4-2}
\usepackage[utf8]{inputenc}
\usepackage[T1]{fontenc}

\usepackage{graphicx}
\usepackage{upgreek}

\usepackage{xcolor}
\usepackage[table]{xcolor}

\usepackage{svg}
\usepackage{textcomp}
\usepackage{booktabs}
\usepackage[normalem]{ulem}
\usepackage{hyperref}
\usepackage{float}
\usepackage{braket}
\usepackage{siunitx}
\preprint{APS/123-QED}
\usepackage{todonotes}

\usepackage{algorithm}
\usepackage{algpseudocode}
\hypersetup{colorlinks=true,
linkcolor=blue,
citecolor=blue,
urlcolor=blue,
filecolor=blue
}

\newcommand{\EVS}{\ensuremath{E_\mathrm{VS} }} 
\newcommand{\vsv}{\Delta_\text{sv}}  

\newcommand{\add}[1]{#1}
\newcommand{\delete}[1]{}

\begin{document}
\title{\delete{Reduction of the impact} \add{Impact} of the local valley splitting on the coherence \\ of conveyor-belt spin shuttling in $^{28}$Si/SiGe}

\author{Mats Volmer}

\affiliation{JARA-FIT Institute for Quantum Information, Forschungszentrum J\"ulich GmbH and RWTH Aachen University, Aachen, Germany}
\author{Tom Struck}
\affiliation{JARA-FIT Institute for Quantum Information, Forschungszentrum J\"ulich GmbH and RWTH Aachen University, Aachen, Germany}
\author{Arnau Sala}
\affiliation{JARA-FIT Institute for Quantum Information, Forschungszentrum J\"ulich GmbH and RWTH Aachen University, Aachen, Germany}
\author{Jhih-Sian Tu}
\author{Stefan Trellenkamp}
\affiliation{Helmholtz Nano Facility (HNF), Forschungszentrum J\"ulich, J\"ulich, Germany}
\author{Davide Degli Esposti}
\author{Giordano Scappucci}
\affiliation{QuTech and Kavli Institute of Nanoscience, Delft University \\ of Technology, Lorentzweg 1, 2628 CJ Delft, The Netherlands}
\author{{\L}ukasz Cywi{\'n}ski}
\affiliation{Institute of Physics, Polish Academy of Sciences, Warsaw, Poland}
\author{Hendrik Bluhm}
\author{Lars R. Schreiber}
\email{lars.schreiber@physik.rwth-aachen.de}
\affiliation{JARA-FIT Institute for Quantum Information, Forschungszentrum J\"ulich GmbH and RWTH Aachen University, Aachen, Germany}

\begin{abstract} 
Electron spins in silicon offer a promising path toward scalable, fault-tolerant quantum computing, with the potential to host millions of qubits. However, scaling up dense quantum-dot arrays and enabling qubit interconnections through shuttling are hindered by uncontrolled lateral variations of the valley splitting energy $\EVS$. We map $\EVS$ across a \SI{40}{\nano\meter} $\times$ \SI{400}{\nano\meter} region of a $^{28}$Si/Si$_{0.7}$Ge$_{0.3}$ shuttle device and analyze the spin coherence of a single electron spin transported by conveyor-belt shuttling. We observe that the $\EVS$ varies over a wide range from \SI{1.5}{\micro\electronvolt} to \SI{200}{\micro\electronvolt} and is dominated by SiGe alloy disorder. In regions of low $\EVS$ and at spin-valley resonances, spin coherence is reduced and its dependence on shuttle velocity matches predictions. Rapid and frequent traversal of low-$\EVS$ regions induces a regime of enhanced spin coherence explained by motional narrowing. By selecting shuttle trajectories that avoid problematic areas on the $\EVS$ map, we achieve transport over tens of microns with coherence limited by the coupling to a static electron spin entangled with the mobile qubit. Our results provide experimental confirmation of the theory of spin-decoherence of mobile electron spin-qubits and present practical strategies to integrate conveyor-mode qubit shuttling into silicon quantum chips.
\end{abstract}
 
\flushbottom
\maketitle

\section*{Introduction}
Single electron spins confined to gate-defined static quantum dots (QDs) in highly isotopically purified planar $^{28}$Si/SiGe quantum wells \cite{Burkard23} represent excellent spin qubits~\cite{Struck2020, philips22, Stano22, defuentes25}. Their manipulation, initialization, and detection fidelities~\cite{Yoneda2018, Struck21, Noiri2022, Mills2022, Xue2022,Wu25} have reached the quantum error correction threshold and are competing with other qubit platforms~\cite{Bluvstein22, Arute19, Monroe21}. Spin qubits in silicon excel due to their tiny footprint and compatibility with silicon foundry fabrication \cite{zwerver22, neyens24, George25, Huckemann25}. Proposals for scalable architectures address cross-talk and wiring fanout bottlenecks \cite{Vandersypen17} by employing mobile spin qubits \cite{Fujita17,Mills19,Yoneda21,noiri22_2,Zwerver23,foster24} and are applicable to quantum error correction protocols \cite{Taylor05, Boter22, Kuenne23, Ginzel24,Yenilen25}.

In particular, the transport of electrons across distances of tens of microns, confined to one of 34 (linear) and 51 (T-junction) moving Si/SiGe quantum dots, has been demonstrated \cite{Xue23, Beer25}. Such a conveyor-mode shuttler, also called QuBus, requires only four control signals, independent of its length \cite{Seidler22, Langrock23}, which is highly advantageous for co-integration with classical control tiles \cite{Vandersypen17, Zhao25}. High-fidelity spin qubit transport, two-qubit gates, and entanglement distribution have also been demonstrated using conveyor-mode shuttling \cite{Struck23,desmet24, matsumoto25}. However, the shuttle lanes in such devices have so far been shorter than 0.5\,µm. According to theory, spin coherence during conveyor-mode shuttling is limited by valley excitations, which depend on shuttle velocity, magnetic field, and valley coupling \cite{Langrock23, Lima24, Lima25}. The latter is determined by the local arrangement of Ge atoms at the alloy-disordered Si/SiGe interface at the position of the quantum dot~\cite{Hollmann20, Wuetz2022, volmer24, Marcks25, Klos24, thayil25}. In particular, uncontrolled excitations might occur at spots with low valley splitting energy $\EVS$, which is the energy difference between the valley ground state and the first excited valley state~\cite{Friesen2007}. Enhancing valley splitting has therefore become crucial for the Si/SiGe platform~\cite{Losert23, woods24, thayil25}, since regions of low $\EVS$  degrade the performance of both static and mobile qubits and pose a substantial challenge for scalability~\cite{Kawakami2014, philips22, Langrock23}. Schemes for optimal control of the spin-qubit shuttling process, balancing fidelity and signal complexity, have been proposed~\cite{Losert24, David24, Nemeth24}, but experimental validation is still lacking. 

Here, we explore this interplay of spin coherence and conveyor-mode shuttling for a known lateral map of $\EVS$ in an isotopically purified $^{28}$Si/SiGe QuBus. First, we employ our shuttle-based method to measure a detailed two-dimensional $\EVS$ map of our device~\cite{volmer24}. Second, we investigate the spin coherence of the shuttle process across a specific path through this $\EVS$ landscape for various magnetic fields, highlighting the impact of spin-valley resonances and regions of low $\EVS$. Finally, we shuttle along optimal and non-optimal trajectories at various velocities and $B$-fields, repeatedly reaching accumulated shuttle distances of several tens of microns with an infidelity of less than 8\% for a shuttle distance of \SI{10}{\micro \meter}.


\begin{figure*}[]
    \centering
    \includegraphics[width=\linewidth]{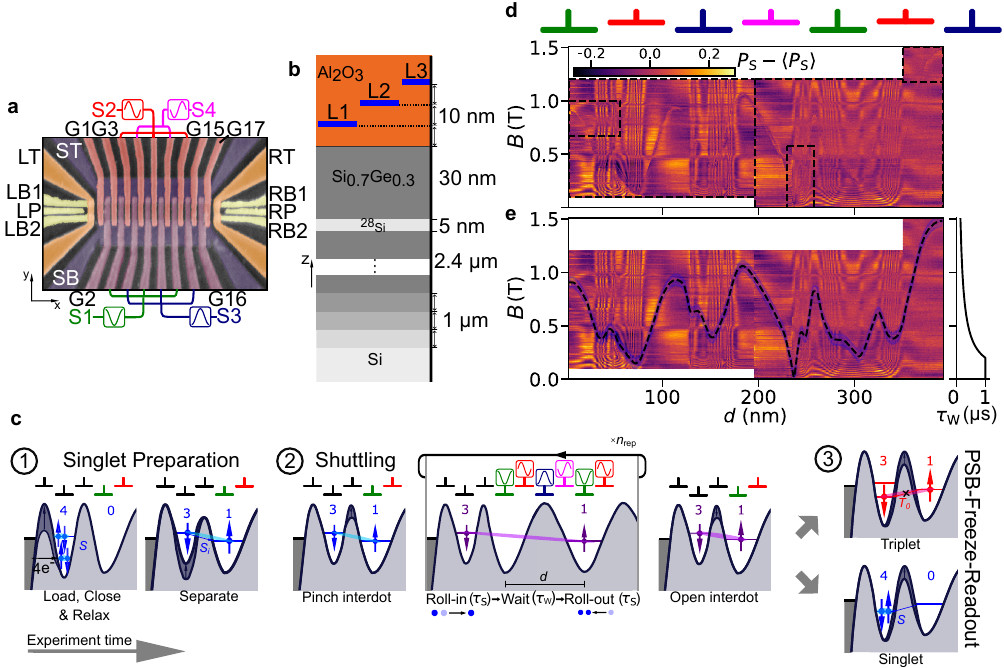}
    \caption{Device and method. (a) False-colored scanning electron micrograph of the three labeled metal layers of the device identical to the one used in this experiment. Electrical connection of S-gates according to solid lines. Scale bar is 500\,nm. (b) Schematic cross-section of the device including Si-buffer, relaxed SiGe (dark gray), Si (white), metallic gates (blue) isolated by Al$_2$O$_3$ (orange). (c) Pulse sequence of the experiment explained by the schematic electrostatic potential at the left side of the one dimensional electron channel (1DEC). Three stages -- singlet preparation, shuttling and Pauli Spin Blockade (PSB) readout are explained in the text. Horizontal lines represent gates and numbers are QD occupation. All pulse sequences are parametrized by \{$d, \tau_\text{W}, \tau_\text{S}, n_\text{rep}, B$\}. (d) Singlet return probability $P_\text{S}$ as a function of magnetic field $B$ and shuttle distance $d$ for $y=\SI{6}{\nano \meter}$ set by constant voltages \SI{0.2}{\volt} and \SI{0.1}{\volt} applied to gates ST and SB, respectively. $P_\text{S}(d, B)$ is composed of five measurements (each enclosed by dashed lines). \add{Positions of clavier gates on L2 and L3 with respect to $d$ are indicated by symbols on top.} (e) Same as d but $\EVS(d)$ is highlighted by the black dashed spline \add{which is carefully positioned by eye}. Right: Wait time $\tau_\text{W}(B)$.}
    \label{fig:Setup_VS_map}
\end{figure*}
We use the left side of a $^{28}\mathrm{Si}$/$\mathrm{Si}_{0.7}\mathrm{Ge}_{0.3}$ linear QuBus shuttle device with two single electron transistors (SETs) at the end of a \SI{1.2}{\micro \meter} long shuttle section defined by three patterned metal layers (Fig.~\ref{fig:Setup_VS_map}a). Shuttle (S)-gates used for the conveyor-mode shuttling in Fig.~\ref{fig:Setup_VS_map}a are electrically connected outside the chip similar to Refs.~\cite{Seidler22,Xue23,Struck23,volmer24}. Thus, the shuttle gates for the four gate sets S1-S4 are controlled by only four voltages. This eases the tuning of the voltages and integration of future cryo-electronics. The SETs are used as proximal charge sensors and electron reservoirs for the linear shuttle section. The device consists of a semiconductor heterostructure as described in Ref.~\cite{paqueletwuetz23} and three electrically isolated metal layers (cross-section in Fig.~\ref{fig:Setup_VS_map}b). Details of the device design and fabrication can be found in the methods section. 

The device is operated at 60\,mK within a global in-plane magnetic field $B$ and controlled by voltage pulses applied to the six gates G2, G3, and S1--S4 (sketched in Fig.~\ref{fig:Setup_VS_map}c). First, the device is tuned electrostatically into the operational regime, at which one SET is formed by gates LB1, LP and LB2, and a double quantum dot (DQD) is formed by gates G1--G3 and the leftmost gates of S1 and S2 (right plunger and right barrier, respectively). Voltages applied to gates S1 through S4 form the QuBus shuttle, which consists of a series of not tunnel-coupled QDs. Each pulse sequence starts with the singlet preparation (1st step in Fig.~\ref{fig:Setup_VS_map}c). First, we (re)load four electrons into the leftmost quantum dot (QD) and wait for \SI{100}{\micro \second} until the loaded spin mixture has relaxed into two spin-singlet states in the two lowest \delete{valley-orbit} \add{orbital
and valley} states available. Then, we tunnel one of the spin-entangled electrons into the right QD of the DQD and raise the interdot barrier by a negative voltage on gate G3 (2nd step in Fig.~\ref{fig:Setup_VS_map}c). Now, the system can be described by a spin-singlet (S) state \delete{with a potentially mixed valley-state}, $S_i$, where $i\in\{\ket{+-},\ket{--},\ket{++},\ket{-+}\}$~\cite{Cywinski25} \delete{represents the valley state of} \add{labels the singly-occupied valley states in} the left and the right QD. 
Two electrons in the left QD remain as an additional, but inert spin-singlet~\cite{volmer24} and can be neglected for further state description. At the third stage (Fig.~\ref{fig:Setup_VS_map}c), the conveyor-mode shuttling is activated by applying sinusoidal voltage pulses  $V_{\text{S},i}$ on the four gates S$_i$
\begin{equation}
    V_{\text{S},i}=A_i\cdot\cos(2\pi f t-(i-1)\cdot\pi/2)+B_i.
\end{equation}
\noindent where $f$ is the drive frequency, $A_i$ is the constant drive amplitude (\SI{150}{\milli \volt} for S1 and S3, \SI{180}{\milli \volt} for S2 and S4) and $B_i$ is the constant voltage offset (\SI{0.7}{\volt} for S1 and S3, \SI{0.84}{\volt} for S2 and S4, similar to \cite{volmer24}). This pulse separates the right electron spin in a moving QD from the other three electrons, which remain in the static leftmost QD formed underneath the gate G2. The corresponding one-way shuttle time $\tau_\text{S}/2$ and distance $d$ are given by $d=\lambda\cdot f\cdot\tau_\text{S}/2$, where $\lambda=\SI{280}{\nano \meter}$ is the period of the shuttle signal (equal to four times the gate pitch), while the shuttling velocity is given by $v_S  \! \equiv \! 2d/\tau_\text{S}$.
After shuttling, we optionally wait for a time $\tau_\text{W}$ before shuttling back by the time-reversed shuttle pulse, thus a shuttle cycle takes $\tau_\text{S}+\tau_\text{W}$. Optionally, the shuttle cycle is repeated $n_\text{rep}$ times, resulting in a total time of $n_\text{rep}(\tau_\text{S}+\tau_\text{W})$. After shuttling, we lower the interdot barrier and detune the DQD into Pauli Spin Blockade (PSB) for \SI{500}{\nano \second} to trigger the spin-selective charge transition. Finally, the charge state is frozen by raising the interdot barrier (3rd step in Fig.~\ref{fig:Setup_VS_map}c) and the spin-singlet (S) is discriminated from the spin-triplet (T) by the SET current in a single shot. The singlet return probability $P_\text{S}$ is then calculated from the statistics of at least 800 pulse sequence repetitions.   

\section*{Results}
\subsection*{Valley splitting map}
In order to explore the impact of $\EVS$ on spin-coherent conveyor-mode shuttling, we map the local valley splitting utilizing the strong renormalization of the shuttled electron $g$-factor $g_{R\nu}$ ($\nu$ represents the valley state) when the local valley splitting matches the Zeeman splitting~\cite{volmer24}. Notably, we directly measure the valley splitting of a QD. \add{The matching is independent of the initialized valley state (see Supplementary Note I)} and we do not approximate the valley splitting by measuring the singlet-triplet splitting within one QD as it is the case for other methods. For one trace $\EVS(d)$, we record $P_\text{S}$ for various $B$ (stepped by \SI{5}{\milli\tesla}) following the  pulse sequence using $n_\text{rep}=1$, a constant frequency of $f=\SI{20}{\mega \hertz}$ for all $d\in[\SI{0}{\nano\meter}, \SI{392}{\nano\meter}]$ (Fig.~\ref{fig:Setup_VS_map}d). In contrast to Ref.~\cite{volmer24}, we observe larger variations of $\EVS$ and need to expand the scanned $B$ range in order to identify the spin-valley anticrossings for all $d$ (black dashed spline line in Fig.~\ref{fig:Setup_VS_map}e), where $\EVS$ equals the local Zeeman energy of the shuttled QD positioned at distance $d$ for a fixed time $\tau_\text{W}(B)$ as plotted in the right panel of Fig.~\ref{fig:Setup_VS_map}e. The chosen dependence of $\tau_\text{W}$ on $B$ ensures that the acquired Larmor-phase  $\varphi_W \propto B\cdot\tau_\text{W}$ at $d$ is balanced with $B$ for optimal visibility of the spin-valley anticrossings (see Supplementary Note II for details).

Indeed, we track $\EVS(d)$ despite some $P_\text{S}$ background due to the static QD \add{(at $B=\SI{0.95}{\tesla}$)} and some vertical features due to a large change of electron g-factor difference $\Delta g_{\mu\nu}(d)=g_{L \mu}-g_{R \nu}(d)$ where $L$ is the static QD and $R$ the mobile QD, while $\mu,\nu \in\{-,+\}$ are the valley indices, see  Ref.~\cite{volmer24}. Additionally, to compensate for slow variations in detection contrast, we align the data by subtracting the linewise mean $\langle P_\text{S} \rangle$. Note that the spin-valley resonances and thus $\EVS(d)$ seamlessly match at the borders of the $P_\text{S}(B,d)$ patches (dashed lines in Fig.~\ref{fig:Setup_VS_map}d) measured on different days. This underlines that the $\EVS(d)$ is a robust and static property of the shuttle device. Sometimes remeasured patches with higher resolution were required if the $\EVS(d)$ trace could not be identified unambiguously in the first place. 

\begin{figure*}
    \centering
    \includegraphics[width=\linewidth]{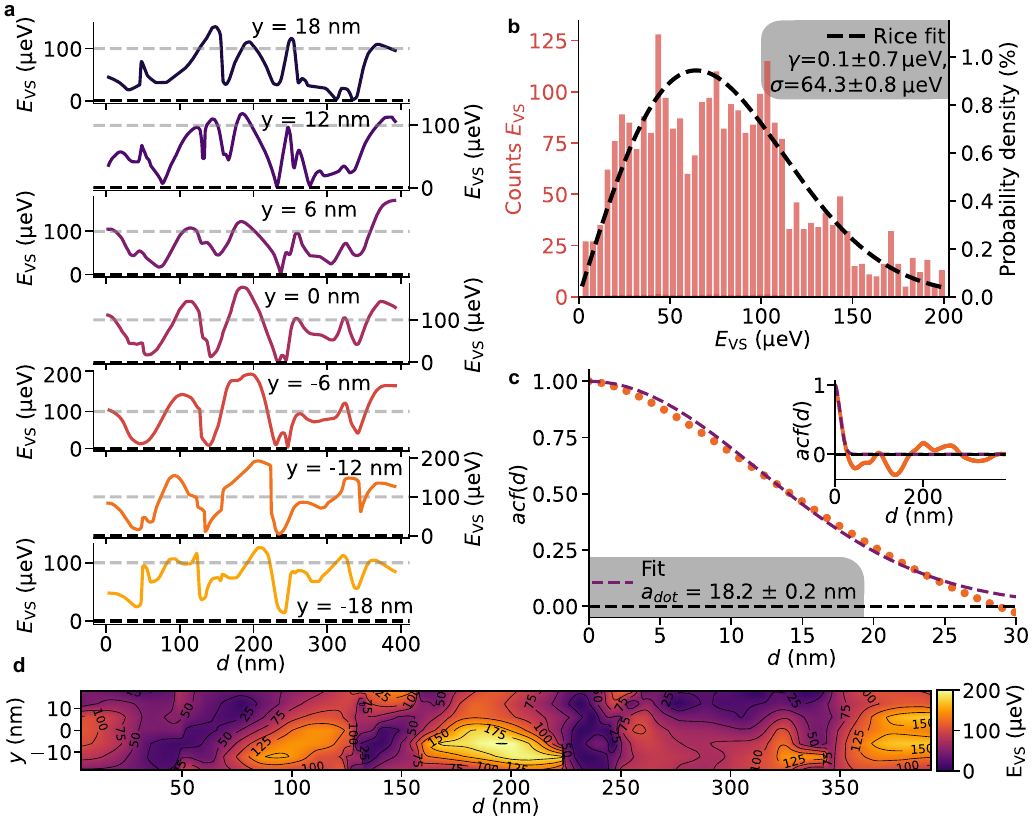}
    \caption{Valley splitting map. (a) Valley splitting traces as a function of shuttle distance $d$ for different 1DEC positions $y$. (b) Histogram of the valley splitting measurements. A fit of a Rice probability density (black dashed line) is parameterized by mean $\gamma$ and width $\sigma$ (cmp. Ref. \cite{volmer24}). (c) Autocorrelation function ($\text{acf}$) of the valley splitting as a function of shuttle distance $d$ (averaged over all traces). A Gaussian fit (dashed, purple) is included. Inset: Zoom-out. (d) Two-dimensional valley splitting map $\EVS(d,y)$ by linear interpolation of the traces in a.}
    \label{fig:VS_map}
\end{figure*}
In order to extend to a two-dimensional map $\EVS(d,y)$, we repeat the measurement protocol above and record seven traces that are offset by \SI{6}{\nano \meter} perpendicular to the \add{one-dimensional electron channel (}1DEC\add{)} (see Fig.~\ref{fig:VS_map}a). Each trace contains 400 measurement points along the \SI{392}{\nano \meter} shuttle distance (see Supplementary Note III for raw data of every trace). The long split gate (gates SB and ST) is set to \SI{150}{\milli \volt} at $y=\SI{0}{\nano \meter}$. The perpendicular offset is set by applying a symmetric voltage bias of \SI{100}{\milli \volt} per \SI{6}{\nano\meter} on gates ST and SB. We calibrated this offset by triangulation of the QDs $y$-position in a similar device~\cite{volmer24}. 

We observe several spots at which $\EVS$ is close to zero with a global minimum of $\EVS=1.5 \pm\SI{1.2}{\micro \electronvolt}$ at $(d,y)=(\SI{231}{\nano \meter},\SI{12}{\nano \meter})$. According to our literature research, this is the smallest $\EVS$ value directly measured and published so far. For example, we find two close local minima with $\EVS \leq \SI{5}{\micro \electronvolt}$ at $(d,y)=(\SI{232}{\nano \meter},\SI{0}{\nano \meter})$ and at $(d,y)=(\SI{245}{\nano \meter},\SI{0}{\nano \meter})$, which merge into one local minimum towards $y=\pm 12$~nm. Such features were observed in a previous two-dimensional map of \EVS~\cite{volmer24} and can be motivated by $\EVS$ being proportional to the modulus of the complex \delete{spin-valley coupling $\vsv$} \add{inter-valley coupling $\Delta$} (see \cite{Wuetz2022} and Supplementary Note IV). Remarkably, the two traces for $y=\pm \SI{18}{\nano \meter}$ exhibit $\EVS>\SI{20}{\micro \electronvolt}$ spanning from $d=\SI{0}{\nano \meter}$ to $d=\SI{280}{\nano \meter}$.
 
From the full two-dimensional $\EVS$ map, we calculate a histogram of $\EVS$ containing 2800 samples (Fig.~\ref{fig:VS_map}b), as a total trace spanning \SI{392}{\nano \meter} contains 400 $\EVS$ points. The histogram follows a Rice distribution~\cite{volmer24} that has a small deterministic component \delete{$\gamma=\SI{0.0\pm0.6}{}$ $\mu$eV and a large spread $\sigma=\SI{61.5\pm0.8}{}$ $\mu$eV} \add{$\gamma=\SI{0.1\pm0.7}{\micro \electronvolt}$ and a large spread $\sigma=\SI{64.3\pm0.8}{\micro \electronvolt}$} suggesting larger impact of alloy disorder than in another heterostructure measured before~\cite{Chen21,volmer24,Marcks25}.
This can be explained by the thinner quantum well, which results in more wave function overlap with the SiGe barrier, leading to larger average $\EVS$ than previously measured \cite{volmer24}, but at the expense of larger variation of valley splitting.

The histogram of the $\EVS$ map thus gives important insight into the origin of $\EVS$ and can be employed as a benchmark for the quality of the semiconductor heterostructure. We also compute the autocorrelation function $\text{acf}(d)$ for all seven traces and plot the average in Fig.~\ref{fig:VS_map}c. Fitting the $\text{acf}$ and thus the size of the shuttled QD by \add{the formula that applies when spatial randomness of $E_{VS}$ is due to alloy disorder at the interface  } \cite{Losert23,volmer24}
\begin{equation}
    \mathrm{acf}(d)=\exp\left(-\frac{1}{4-\uppi} \frac{d^2}{a_\mathrm{dot}^2}\right),
    \label{eq:corr}
\end{equation}

\noindent we find a QD \delete{size}\add{radius} of \delete{$a_\text{dot}=\SI{17.3\pm0.1}{\nano \meter}$}\add{$a_\text{dot}=\SI{18.2\pm0.2}{\nano \meter}$}, similar to Ref.~\cite{volmer24} \add{(\SI{16}{\nano \meter}) and Ref.~\cite{Marcks25} (\SI{19.2}{\nano \meter})}. The inset shows no correlation beyond \SI{30}{\nano \meter}. This implies that samples in the histogram are correlated and deviations from the Rice distribution predicted by theory of SiGe alloy disorder are not significant. Finally, we combine all $\EVS$ traces by linear interpolation to create one two-dimensional valley splitting map in Fig.~\ref{fig:VS_map}d. This map extends to almost four times the size of a previously published map of a heterostructure with natural abundance of silicon isotopes~\cite{volmer24} and has a three times larger span of valley splitting values. \add{This implies 122 samples of measured $\EVS$ spaced by a distance of a QD radius being the correlation length.} As the mapping method might become a significant tool for tracking the material progress for Si/SiGe qubit chips, we added comments on the time efficiency of the method into the method section Efficiency of valley mapping.

\subsection*{Spin decoherence during shuttling}

\begin{figure*}
    \centering
    \includegraphics[width=\linewidth]{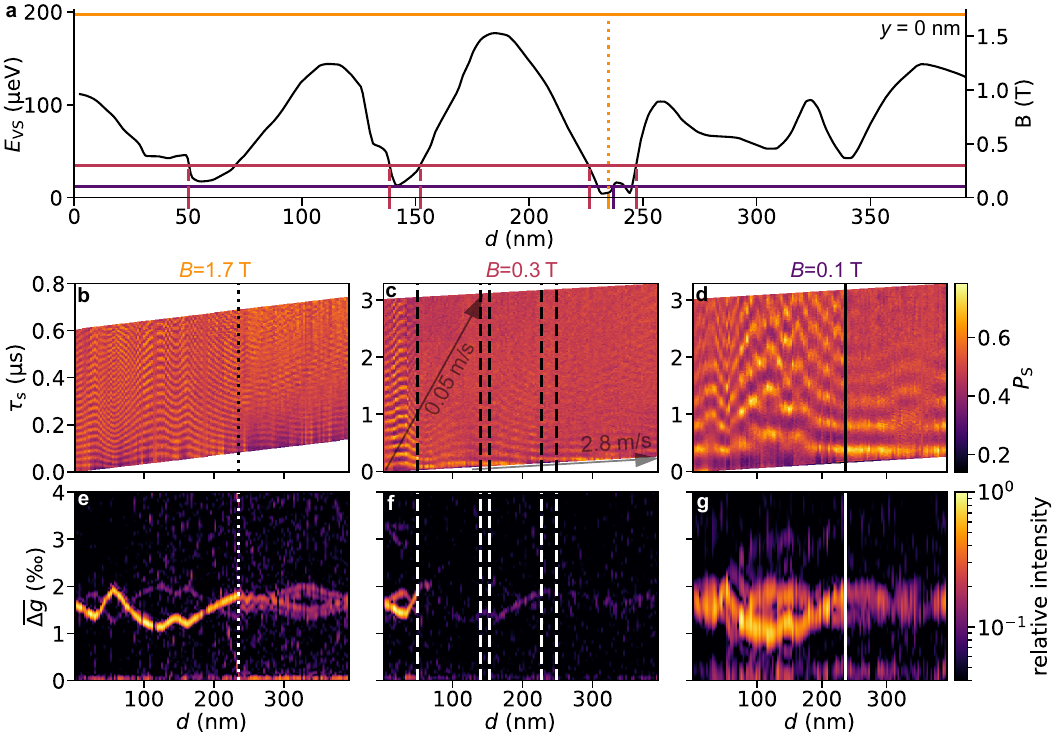}
    \caption{Impact of $\EVS$ on coherent spin shuttling. (a) Valley splitting trace at $y=\SI{0}{\nano \meter}$ from Fig.~\ref{fig:VS_map}a with colored horizontal lines indicating the Zeeman energy for $B \in \{1.7, 0.3, 0.1\}$\,T. Spin-valley resonances are marked by vertical dashed lines and a region of low $\EVS$ by a dotted line. (b,c,d) Singlet-return probability $P_\text{S}$ as a function of shuttle distance $d$ and shuttle time $\tau_\text{S}$ (for one direction) measured with the fixed three magnetic fields $B \in \{1.7, 0.3, 0.1\}$\,T are shown in b,c and d, respectively. Vertical lines indicate spin-valley resonances and region of low $\EVS$ from a. Traces for fixed shuttle velocities (e,f,g) are indicated by the arrows in c. (e,f,g) FFT along $\tau_\text{S}$ of the data shown above as a function of integrated g-factor difference $\overline{\Delta g}$ for magnetic fields $B \in \{1.7, 0.3, 0.1\}$\,T. \add{ A version of panels (b-g) can be found in Supplementary Note V.}}
    \label{fig:shuttle_time_measurements}
\end{figure*}
Next, we use the knowledge of the local valley splitting landscape to explore the impact of local $\EVS$ on the spin coherence during conveyor-mode shuttling. We start with a shuttle trajectory set to $y=\SI{0}{\nano \meter}$ with $\EVS(d,y=0)$ shown in Fig.~\ref{fig:shuttle_time_measurements}a. We explore $P_\text{S}(d, \tau_\text{S})$ for a fixed magnetic field $B$ and $n_\text{rep}=1, \tau_\text{W}=0$.

We start with $B=\SI{1.7}{\tesla}$ (Fig.~\ref{fig:shuttle_time_measurements}b), since \add{the} Zeeman energy of the moving QD satisfies $E_\text{Z} > \max(\EVS(d))$ (yellow line in Fig.~\ref{fig:shuttle_time_measurements}a) and we expect no \delete{spin decoherence} \add{drop in ST-oscillation amplitude} at spin-valley resonances defined by $E_\text{Z} = \EVS$ \cite{Langrock23}. For $P_\text{S}(d, \tau_\text{S})$, we use various shuttle velocities, and some $P_\text{S}(d, \tau_\text{S})$ are inaccessible (white areas) due to the maximum shuttle velocity of \SI{5.6}{\meter \per \second}. Similar to Ref.~\cite{Struck23}, we observe variations in the singlet-triplet oscillation frequency stemming from $\Delta g(d)$. More importantly, the amplitude of the singlet-triplet oscillations \delete{and thus the spin coherence} is suddenly reduced at $d \approx \SI{240}{\nano \meter}$ (dotted vertical lines in Fig.~\ref{fig:shuttle_time_measurements}a,b). This is remarkable and not unexpected, as this sudden loss of \delete{spin coherence} \add{ST-oscillation amplitude} coincides with the first spot at which $\EVS \! < \! \SI{5}{\micro \electronvolt}$. 

The $P_\text{S}(d, \tau_\text{S})$ measured at $B=\SI{0.3}{\tesla}$ (Fig.~\ref{fig:shuttle_time_measurements}c with maximal velocity \SI{2.8}{\meter \per \second}) exhibits  a smaller frequency of ST-oscillations, as expected, but also reveals very different \delete{spin-decoherence} \add{signal-decay} characteristics. Here, we observe a sudden reduction of the visibility of the oscillations and thus increase of \delete{spin-decoherence} \add{signal decay} after passage beyond $d=\SI{50}{\nano \meter}$ (dashed line in Fig.~\ref{fig:shuttle_time_measurements}c). Remarkably, this coincides with the first spin-valley resonance passed by shuttling (dashed vertical red lines in Fig.~\ref{fig:shuttle_time_measurements}a). Repeating the measurement with a reduced $B=\SI{0.1}{\tesla}$ (Fig.~\ref{fig:shuttle_time_measurements}d), the sudden loss of \delete{spin coherence} \add{signal} appears at larger $d=\SI{240}{\nano \meter}$ (black solid line), which coincides with the first spin-valley resonance at the corresponding reduced $E_\text{Z}$ (solid violet line in Fig.~\ref{fig:shuttle_time_measurements}d). Furthermore, when following $P_\text{S}(d, \tau_\text{S})$ oscillations for constant shuttle velocities (see arrows in Fig.~\ref{fig:shuttle_time_measurements}c as a guide),  we observe that shuttling across spin-valley resonances with higher velocity tends to preserve the \delete{spin coherence} \add{ST-oscillations} compared to lower shuttle velocities for both magnetic fields. Specifically, at the largest $v_\text{S}=\SI{2.8}{\meter /\second}$, the oscillation amplitude is unchanged and the \delete{spin coherence} \add{signal strength} is not affected by the resonance. The opposite trend is observed if the onset of \delete{spin decoherence} \add{signal decay} coincides with the crossing of valley splitting minimum in Fig.~\ref{fig:shuttle_time_measurements}b.   

We support our observations by the Fourier transforms of $P_\text{S}(d, \tau_\text{S})$, plotted as functions of frequency $f$ divided by $2\mu_B B/h$ (Fig.~\ref{fig:shuttle_time_measurements}e-g) underneath the corresponding panels Fig.~\ref{fig:shuttle_time_measurements}b,c,d. This extracts the integrated difference $\Delta g$ of the electron $g$-factors for various shuttle distances $d$, because the acquired phase during shuttling is the integral over the Larmor phases $\phi$ the electron spin acquires along its shuttle trajectory \cite{Struck23}
\begin{equation}
    \phi_{\mu\nu}=2\cdot\int_0^{\tau_\text{S}/2}  \frac{\mu_B B}{\hbar}\Delta g_{\mu\nu}(x(t))dt\approx\frac{\mu_B B}{\hbar} \overline{\Delta g}_{\mu\nu}(d) \cdot \tau_\text{S},
    \label{eq:integral}
\end{equation}
\noindent where $\overline{\Delta g}_{\mu\nu}(d) =\frac{1}{d}\int_{0}^{d} \Delta g_{\mu\nu}(x)\mathrm{d}x$, and we have assumed that $v_\text{S}$ is approximately constant despite the presence of electrostatic disorder. 

In Fig.~\ref{fig:shuttle_time_measurements}e, we observe one dominant $\overline{\Delta g}_{\mu\nu}(d)$ and one faint component amplified by the log-color-scale. They correspond to two distinct singlet-triplet frequencies. Importantly, the amplitude of both drastically change at approximately \SI{240}{\nano \meter}, matching the minimum of $\EVS$, at which a sudden reduction of the spin coherence is observed. Strikingly, the two frequency components split into four. We interpret these frequencies in terms of mixtures of valley states of the initialized spin-singlet: Before shuttling the singlet is in a dominant valley state and a small admixture of a different valley state in the static QD \cite{Cywinski25}. This remains unchanged during the shuttling through regions of large $\EVS$. After passing the $\EVS$ minimum, however, both valley states of the shuttled QD become occupied as well (see Supplementary Notes IV and VI), resulting in a total of four singlet-triplet frequencies corresponding to four $\mu\nu \in \{++,--,+-,-+\}$ combinations. Hence, the passage of the $\EVS$ minimum causes a partial loss of spin coherence and partial valley excitation, and therefore the shuttling is not fully adiabatic \cite{Langrock23}. The magnitude of the $\overline{\Delta g}_{\mu\nu}(d)$ components and their symmetry will be explained in more detail elsewhere, but it is important to keep in mind that $\overline{\Delta g}_{\mu\nu}(d)$ depends on all $\Delta g_{\mu\nu}(x)$ on the shuttle trajectory according to Eq. \ref{eq:integral}.

\begin{figure*}
    \centering
    \includegraphics[width=\linewidth]{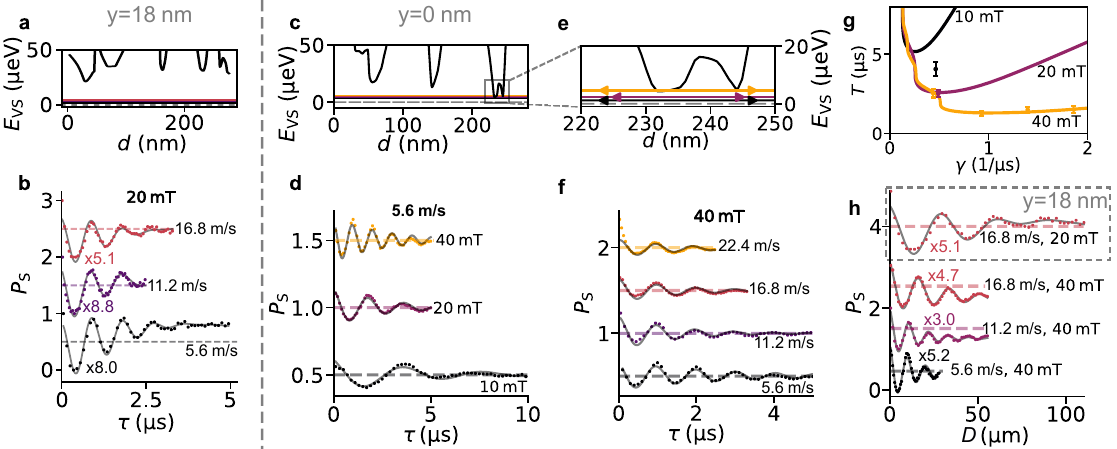}
    \caption{Repetitive impact of shuttling through an $\EVS(x)$ landscape on coherent spin shuttling. (a) Zoom-in of $\EVS$ trace at $y=\SI{18}{\nano \meter}$ from Fig.~\ref{fig:VS_map}a with colored horizontal lines indicating the Zeeman energy for $B=20$\,mT. (b) $P_\text{S}$ as a function of total shuttle time $\tau=2n_\text{rep}\tau_\text{S}$ recorded at $B=20$\,mT and at labeled shuttle velocity $v_\text{S}$. Every data point represents $n_\text{rep}$ shuttling of distance \SI{280}{\nano\meter} and back at $y=\SI{18}{\nano \meter}$. Solid gray lines are least-squares fits to the data. Dashed horizontal lines are the mean of all data points in the scan. Data points are scaled by given factor to normalized visibility to 1. $P_\text{S}$ is offset vertically by 1 for clarity. (c) Zoom-in of valley splitting trace at $y=\SI{0}{\nano \meter}$ from Fig.~\ref{fig:VS_map}a with colored horizontal lines indicating the Zeeman energy for $B \in \{10, 20, 40\}$\,mT. (d) Same as in b, but $y=\SI{0}{\nano \meter}$ and fixed shuttle velocity $v_\text{S}=\SI{5.6}{\meter / \second}$ and $B \in \{10, 20, 40\}$\,mT. $P_\text{S}$ is offset vertically by 0.5 for clarity. (e) Same as in c, but larger zoom. (f) Same as in d, but for fixed magnetic field $B=40$\,mT and shuttle velocities $v_\text{S}\in \{5.6, 11.2, 16.8, 22.4\}\SI{}{\meter / \second}$. For the fit of the \SI{22.4}{\meter / \second} dataset, we exclude the first three points for fitting.  (g) Modeled exponential decay \add{timescale} $T$ for repetitive impact of $\EVS$ minimum as a function of valley excitation rate $\gamma$ and $B$ (solid lines) with the six fitted $T$ values (see also Tab.~\ref{tab:Fit_params_cum_shut}) from panels d and f marked as solid circles. (h) $P_\text{S}$ as a function of total shuttle distance $D=2n_\text{rep}\lambda$ for various $(v_\text{S},B,y)$. Data and fits are taken from panels b,d,f and rescaled according to label with $P_\text{S}$ offset vertically for clarity.}
    \label{fig:cumulative_shuttling}
\end{figure*}

The Fourier transform of the singlet-triplet data recorded at $B=\SI{0.3}{\tesla}$ (Fig.~\ref{fig:shuttle_time_measurements}f) again reveals the initialization of a mixture of valley occupations, with one component dominant over the other. After the passage of the first spin-valley resonance, both components nearly vanish. Interestingly, at larger $d$ the visibility of one component tends to recover and vanish in the vicinity of the passage through the next resonance marked by another vertical dashed line (Fig.~\ref{fig:shuttle_time_measurements}f, $d=\SI{150}{\nano \meter}-\SI{220}{\nano \meter}$). The reason for the apparent recovery of the oscillations is a peculiarity of the $P_\text{S}(d, \tau_\text{S})$ measurement scheme. Larger values of $d$ are dominantly recorded at larger shuttle velocities, and consequently shuttle sequences with small velocities do not contribute to the Fourier transform at larger $d$. Since the passage of spin-valley resonances with larger velocities tends to preserve the spin coherence, as we noted above, the coherence at larger $d$ tends to recover. A natural explanation of this behavior is the transition between adiabatic and diabatic passage through the spin-valley resonance \cite{Langrock23}. The adiabatic passage at low $v_\text{S}$ leads to spin-valley flip-flop that converts superposition of spin states into a superposition of valley states (see Supplementary Note VII). The latter is then rapidly dephased due to valley splitting fluctuations (see Supplementary Note VIIIB).
We quantitatively confirm this effect by simulation of the data in Fig.~\ref{fig:shuttle_time_measurements}f based on realistic assumption of the spin-valley-coupling $\Delta_\text{sv} \lesssim 300$\,neV and the recorded $\EVS$ map (see Supplementary Note IX), from which $\mathrm{d}\EVS/\mathrm{d}x \approx  \SI{3}{\micro\electronvolt\per\nano\meter}$ is extracted at the resonance. The probability $P_{\text{svf}}$ for a spin-valley flip-flop at the spin-valley resonance is:
\begin{equation}
P_{\text{svf}} = 1- \exp\left(-2\pi \vsv^2 / \hbar v_\delta\right), \label{eq:Psvf}
\end{equation}
\noindent where $v_\delta \! \equiv \! |\mathrm{d}\EVS/\mathrm{d}x| \cdot v_\text{S}$. For $v_\text{S}>\SI{2.8}{\meter /\second}$, we find 
$P_{\text{svf}} \! \ll \! 1$;
thus the passage is diabatic, and the spin-valley flip-flop is suppressed.

The Fourier transform of the singlet-triplet data recorded at $B=\SI{0.1}{\tesla}$ is broadened as only a few singlet-triplet oscillations are recorded (Fig.~\ref{fig:shuttle_time_measurements}g). Most importantly, the amplitudes of the components abruptly diminish at the passage of the first spin-valley resonance located at $d=\SI{230}{\nano \meter}$ according to the mapped $\EVS$. Hence, it confirms our notion of the impact of the spin-valley resonance for the spin coherence of the shuttle process.

In summary, we observe two spin dephasing channels directly related to the $\EVS$ map. First, passing a region of small $\EVS$, the spin coherence is partly lost and both valley states in the mobile QD become occupied. Second, passing a spin-valley resonance, defined by $E_\text{Z}=\EVS$, with low $v_\text{S}$ results in conversion of spin qubit to a valley qubit that suffers stronger dephasing caused by $\EVS$ fluctuations. At higher velocities (here $v_\text{S}>\SI{2.8}{ \meter /\second}$), the passage can be fully diabatic and no spin-valley-flip occurs. In Supplementary Note VIII, all mechanisms of decoherence during the shuttling such as hyperfine noise (VIIIA), valley splitting fluctuations activated by spin-valley flip-flop (VIIIB), spin relaxation near a spin-valley resonance (VIIIC) and valley excitation (VIIID) are discussed in more details. Calculation reproducing the features seen in Fig.~3f is described in Supplementary Note IX.

\subsection*{Shuttling across larger distances}
Finally, we quantitatively investigate the spin coherence of the shuttle process in an extended parameter space. In order to amplify the impact of spots of interest on the $\EVS$ map, we repeat the shuttle process forward and backward $n_\text{rep}$ times at a fixed shuttle trajectory of distance $\lambda=\SI{280}{\nano\meter}$ (one period), a fixed shuttle velocity $v_\text{S}=\lambda f$, a fixed magnetic field $B$ and $\tau_\text{W}=0$ (Fig.~\ref{fig:Setup_VS_map}c). Then the decay of $P_\text{S}$ (the loss of coherence of the spin singlet) as a function of the accumulated shuttle time $\tau$ at a total distance $D=2 n_\text{rep}\cdot\SI{280}{\nano\meter}$ is fitted. \add{Each measurement was taken over the span of approximately ten minutes.}
 
First, we select a shuttle trajectory without near-zero $\EVS$ values ($y=\SI{18}{\nano \meter}$, $x=0-\SI{280}{\nano \meter}$ in Fig~\ref{fig:VS_map}a) and $B=\SI{20}{\milli \tesla}$ as a reference. Any spin-valley resonances are avoided, since $E_\text{Z} \ll \EVS(x)$ (Fig.~\ref{fig:cumulative_shuttling}a). $P_\text{S}(\tau)$ is measured for three shuttle velocities (Fig.~\ref{fig:cumulative_shuttling}b). We observe decaying spin-singlet oscillations on top of a rising background. Most importantly, the curves and more specifically the decay is independent of the shuttle velocity $v_\text{S}$. Hence, the total shuttle time $\tau$, and not the accumulated shuttled distance $D$ seem to influence the spin decoherence during the shuttle process. Note that in Ref. \cite{Langrock23}, we observed rising $T_2^{*}(d)$ as a function of $d$ due to motional narrowing. Here, motional narrowing might enhance the $T_2^{*}(D)$ as well, but equally for all $D = 2 n_\text{rep} \lambda$, since quasistatic noise is averaged across the distance of $\lambda$ only, but for all $n_\text{rep}$, thus for all data points recorded, see Supplementary Note VIIIA. The rising background might be due to a very slow singlet-triplet oscillation, stemming from a different valley-combination of the occupied singlet state. We capture it by a second frequency component:

\begin{align}
P_\text{S}(\tau,T_2^{*}) &=
  A_1 \cos(\omega_1 \tau + \phi_1)\, e^{-\tau^{2}/T_2^{*2}} \nonumber \\
&\quad + A_2 \cos(\omega_2 \tau + \phi_2) + \tfrac12 + \epsilon, 
\label{eq:ps_gauss_decay}
\end{align}

\noindent where $T_2^{*}$ is the decay of the entangled singlet state and $\epsilon$, $A_i$, $\phi_i$ are constants capturing SPAM errors and relative valley occupation and oscillation phase. We find a good fit with a Gaussian decay, but the first $P_\text{S}$ data point of all three traces is systematically too large for unknown reason. We find $T_2^{*} \approx \SI{1.7}{\micro \second}$ (Tab.~\ref{tab:Fit_params_cum_shut}), which is a reasonable value for a spin singlet decay in an isotopically purified $^{28}$Si/SiGe, since the value is expected to be by a factor of $\sqrt{2}$ smaller than the single spin ensemble dephasing time \cite{Wu25}. Remarkably, the spin-singlet decay in a DQD with raised barrier (shuttle distance $d=0$\,nm) is found to be \SI{1.4}{\micro \second}, which is longer than for the natural Si/SiGe ($\approx \SI{0.6}{\micro \second}$ in Ref. \cite{Struck23}), but shorter than the decay including the shuttle process extended by motional narrowing and presumably limited by the spin dephasing in the static (left) QD. Hence, the shuttle process tends to not add to the spin-decoherence of the spin singlet, if the impact of $\EVS$ minima and spin-valley resonances can be avoided. The shuttle distance can be increased by larger shuttle velocity as only the shuttling time compared to the $T_2^{*}$ corrected by motional narrowing is governing the singlet decay.        

Next, we alter the shuttle trajectory to $y=\SI{0}{\nano \meter}$, $x=0-\SI{280}{\nano \meter}$ and thus cross a region of near-zero $\EVS$ (Fig.~\ref{fig:cumulative_shuttling}c), at which we expect an impact on the spin coherence. We measure $P_\text{S}(\tau)$ for $B \in [10,20,40]$\,mT at a fixed $v_\text{S}=\SI{5.6}{\meter/\second}$ (Fig.~\ref{fig:cumulative_shuttling}d). The trajectory does not cross a spin-valley resonance at $B=\SI{10}{\milli \tesla}$, while there are several such passages at $B=\SI{40}{\milli \tesla}$ (Fig.~\ref{fig:cumulative_shuttling}e). We observe that the decay of $P_\text{S}(\tau)$ is significantly longer for $B=\SI{10}{\milli \tesla}$ and about equal at $B=\SI{20}{\milli \tesla}$ and $B=\SI{40}{\milli \tesla}$. Hence, we can shuttle the spin the longest distance $D$ at $B=\SI{10}{\milli \tesla}$. At first sight, it is surprising that the decay at $B=\SI{20}{\milli \tesla}$ and $B=\SI{40}{\milli \tesla}$ is about equal, as there is no passage of spin-valley resonance at $B=\SI{20}{\milli \tesla}$. Furthermore, we observe that the decay of $P_\text{S}(\tau)$ and thus the spin coherence recorded at $B=\SI{40}{\milli \tesla}$ tends to decrease with increasing shuttle velocity (Fig.~\ref{fig:cumulative_shuttling}f), although $v_\text{S} > \SI{2.8}{\meter/\second}$, and therefore passage across the spin-valley resonances should be almost diabatic (no spin-valley-flip occurs), as we noted above, and the timescale of decay of coherence due to spin-valley flip-flop should be velocity-independent, see Supplementary Note XA.

Therefore, a detailed analysis of the $B$ and $ v_\text{S}$ dependence of the passage of the region with near-zero $\EVS$ is required. We show that the phenomena in Fig.~\ref{fig:cumulative_shuttling}d,f can be explained by the passage of the region of small $\EVS$ alone. First, we note that the form of $P_\text{S}$ decay is better approximated by an exponential than by the Gaussian used in Eq.~(\ref{eq:ps_gauss_decay}) if the decoherence is caused by accumulation of errors occurring over time $\tau$

\begin{align}
P_\text{S}(\tau) &=
   A \cos(\omega \tau + \phi) e^{-\tau/T}
  + \tfrac12 + \epsilon,
\label{eq:ps_exp_decay}
\end{align}

\noindent where $T$ is the fitted decay time (Tab.~\ref{tab:Fit_params_cum_shut}) and $A, \epsilon$ are constants capturing SPAM errors. Indeed, for all parameters $B$ and $v_\text{S}$, except for the lowest values ($B =  \SI{10}{\milli \tesla}$, $v_\text{S} = \SI{5.6}{\meter\per\second}$), the exponential decay fitted the measured data in Fig.~\ref{fig:cumulative_shuttling}d,f significantly better than a Gaussian decay. The first three data points for Fig.~\ref{fig:cumulative_shuttling}f (\SI{22.4}{\meter \per \second}) are excluded from the fit, as they show an additional exponential background. This suggests that for the $y=0$ shuttling trajectory the decoherence for $B >  \SI{10}{\milli \tesla}$ is dominated by dynamic processes affecting the shuttled spin instead of just inhomogeneous spin dephasing $T_{2}^*$.

\begin{table}[t]
  \centering
  \setlength{\tabcolsep}{0.75em}
  \begin{tabular}{ccccc}
    \toprule
    $y$ (nm) & $v$ (\si{\meter \per \second}) & $B$ (mT) &
      $T_2^*$ (\si{\micro\second}) &
      $T$ (\si{\micro\second}) \\
    \midrule

    18 & 16.8 & 20 & 1.5 ± 0.1   & {} \\
    18 & 11.2 & 20 & 1.5 ± 0.1   & {} \\
    18 & 5.6 & 20 & 2.0 ± 0.1   & {} \\
    \midrule
     0 & 22.4 & 40 & {}   & 1.5 ± 0.2 \\
     0 & 16.8 & 40 & {}   & 1.5 ± 0.2 \\
     0 & 11.2 & 40 & {}   & 1.3 ± 0.2 \\
     0 & 5.6 & 40 & {}   & 2.6 ± 0.1 \\
     0 & 5.6 & 20 & {}   & 2.5 ± 0.1 \\
     0 & 5.6 & 10 & {}   & 4.1 ± 0.5 \\
    \bottomrule
  \end{tabular}
  \caption{Fit parameters for the cumulative shuttling measurements. For $y=\SI{18}{\nano \meter}$, we fit the data using Eq.~\ref{eq:ps_gauss_decay}. The $T_2^*$ corresponds to the dephasing parameter in the fit equation. For $y=\SI{0}{\nano \meter}$, we fit the data using Eq.~\ref{eq:ps_exp_decay} and extract the decay time $T$. Note that we measure the dephasing of a separated entangled spin state, and therefore $T_2^*$ and $T$ include dephasing of the spins in the static and the mobile QDs.} 
  \label{tab:Fit_params_cum_shut}
\end{table}

Spin dephasing due to multiple passages through a valley splitting minimum is caused by valley-flip events that lead to accumulation of stochastic phase contributions due to valley-dependent $g$-factor, $g_{R\nu}(d)$ of the shuttled spin, and random times spent in each valley state during the total shuttling time $\tau  = n_\text{rep}\tau_\text{S}$. As valley relaxation can be neglected \cite{Langrock23}, this dephasing is controlled by two parameters: $\delta\bar{\omega}  \equiv  \bar{\omega}_{+}-\bar{\omega}_{-}$, in which $\bar{\omega}_{\nu} = \mu_{B} B \overline{\Delta g}_{\mu\nu}(\lambda)/\hbar$, with $\overline{\Delta g}_{\mu\nu}$ from Eq.~(\ref{eq:integral}), is evaluated for the shuttled spin in valley $\nu  = \pm$, and the effective valley flip rate $\gamma \equiv   Q_v(v_\text{S}) v_\text{S}/\lambda$, in which $Q_v(v_\text{S})$ is the probability of a valley-flip per passage. For $\gamma  \gg  \delta\bar{\omega}$ the net random phase accumulated due to multiple valley flips undergoes motional narrowing, the coherence decay is exponential,  and its timescale is $T\!  =\! 2\gamma/\delta\bar{\omega}^2$. On the other hand, for $\gamma \! \ll \!  \delta\bar{\omega}$ the decay is approximately exponential, with timescale  $T \!= \! 1/\gamma \!  \propto \! 1/(Q_v v_\text{S})$. Thus, the $T(\gamma)$ dependence is non-monotonic, with the minimum dephasing time $T_\text{min} \! \approx \! 2/\delta\bar{\omega}$ obtained for $\gamma  \approx  \delta\bar{\omega}/\sqrt{2}$ (Fig.~\ref{fig:cumulative_shuttling}g). If we assume approximately velocity-independent $Q_{v}$, which holds for strongly nonadiabatic dynamics expected for shuttling in the relevant velocity range for $\EVS(d)$ from Fig.~\ref{fig:cumulative_shuttling}e, 
then $T(\gamma, B)$ depends on only the two QuBus-specific parameters $(Q_{v}, \delta\bar{\omega})$. Fitting $T(B,\gamma)$ from Tab.~\ref{tab:Fit_params_cum_shut} to the theory, we find reasonable values $Q_v  =  0.023\pm 0.003$  and $ \delta\bar{\omega}/\mu_{B}B  \approx  4.8\cdot 10^{-4}\pm 3\cdot 10^{-5}$ yielding a good fit to our model (detailed discussion in Supplementary Note XB). The largest value of $T$ from data recorded at the lowest $B$ is most probably dominated by the dephasing in the static QD, and thus does not reach the theoretical value for the mobile QD. Therefore, we excluded it from the fit of the model. Let us stress the key observation that the decoherence time $T$ can be enhanced not only by lowering $Q_v$ (by lowering $v_\text{S}$, since at low enough velocities $Q_v$ certainly decreases with decreasing $v_\text{S}$) \cite{Langrock23}, but also by increasing $v_\text{S}$, and thus $Q_v$, in order to reach the motional narrowing regime, which is analogous to the magnetic-field-dependent Dyakonov-Perel spin-dephasing mechanisms for carrier-scattering of free spin ensembles \cite{Dyakonov72}. 

Finally, we compare the decay of $P_\text{S}$ in terms of total shuttle distance $D$ for various shuttle trajectories and parameters $v_\text{S}$ and $B$ (Fig.~\ref{fig:cumulative_shuttling}h). Targeting maximum spin-coherent shuttle distance without valley excitations, the shuttle velocity should be high and the trajectory should avoid minima of $\EVS$. Picking such a trajectory, we find a Gaussian decay dominated by the inhomogenous spin-dephasing of the spin in the static QD and enhanced spin-dephasing of the mobile QD due to motional narrowing. For a pessimistic estimate, we assume both ensemble spin-dephasing times to be equal and arrive at a spin shuttle fidelity of $\mathcal{F}=92$\,\% across a total distance $D=\SI{10}{\micro \meter}$ from 
\begin{equation}
    \mathcal{F}(D)=\exp\left(-\left(\frac{D}{v_\text{S}\sqrt{2}T_2^*}\right)^2\right),
\end{equation}
for a shuttling velocity of $v_\text{S}=\SI{16.8}{\meter \per \second}$ at $y=\SI{18}{\nano \meter}$ with $T_2^*=\SI{1.5}{\micro \second}$ for the spin singlet. \add{Comparable literature focused on high-fidelity spin shuttling~\cite{desmet24}, surpasses this. As we did not optimize the experiment for fast shuttle speeds, and have additional decay due to the static electron, this is expected.}

\section*{Discussion}
We have explored the spin decoherence mechanisms during conveyor-mode shuttling of a single electron spin in an isotopically purified $^{28}$Si/SiGe QuBus and its correlation to the measured two dimensional map of valley splitting energies. The values of $\EVS$ in the map range between $\SI{1.5}{\micro \electronvolt}$ and $\SI{200}{\micro \electronvolt}$, and follow predictions of $\EVS$ governed by SiGe alloy disorder \cite{Losert23}. Specifically, the trend towards a larger spread of $\EVS$ with larger mean values matches the thinner quantum well mapped here compared to Ref.~\cite{volmer24}. Controlling the shuttle trajectory, velocity, and magnetic field, we observe the local impact of $\EVS$ during shuttling of a single spin. The observed impact on spin coherence of special points on the $\EVS$ map can be categorized into two types: (I) Decoherence due to valley excitation in the mobile QD. This valley excitation occurs in regions of low $\EVS$. (II) Spin-valley flip-flop due to a passage through a spin-valley resonance that turns the spin superposition into a quickly dephasing valley superposition. Our observations confirm the relevance of these spin-dephasing mechanisms during the conveyor-belt spin shuttling~\cite{Langrock23}, but also trigger the prediction of a new regime of enhanced spin coherence due to a large number of excitation events controlled by enhancing the shuttle velocity. This transition is an analogue of the cross-over of the Elliott-Yafet to the magnetic-field-dependent Dyakonov-Perel spin-dephasing mechanisms \delete{found  at} \add{for} carrier-scattering of free spin ensembles in spintronics \cite{Zutic04}. \add{Additionally mapping the $g$-factor, we can evaluate the complex intervalley coupling, offering a complete analysis of the valley physics}~\cite{volmer26}.
Our work suggests and supports strategies for achieving high-fidelity spin-shuttling at large distances~\cite{Lima23,Oda24,Nemeth24,Pazhedath24}. First, local spots of valley excitations (predominantly found at low $\EVS$) and spin-valley resonances can be avoided by appropriate two-dimensional trajectories~\cite{Losert24}, and/or the overall $\EVS$  has to be enhanced by advanced heterostructures~\cite{McJunkin21,McJunkin22,Losert23,Gradwohl25}. Our mapping method of $\EVS$ is the key to exploit these strategies \add{and we exploit that the $\EVS$ is highly reproducible for a device within a thermal cycle}. We observe that high shuttle velocities are preferable beyond running the quantum chip at a high clock rate: (I) Larger shuttle distances are achieved at a given static $T_2^*$, which is enhanced by motional narrowing of quasistatic spin splitting noise. Here, we find a maximal infidelity of 8\,\% for a shuttle distance of \SI{10}{\micro \meter}. (II) Spots of spin-valley resonance are diabatically passed. At very high shuttle velocities, low magnetic fields and large distances, the Dyakonov-Perel dephasing regime is entered, at which trajectories with many spots of valley excitation are preferred. In our experiment, we trigger this transition by repeatedly passing an $\EVS$ region with two small minima. Notably, our measurements discriminate only spin states and probably shuttling in the Dyakonov-Perel regime results in a mixed occupancy of the involved two valley states in the mobile QD. This poses a problem for spin manipulation unless mitigated by valley relaxation. 

The transition to mobile spin qubits and high-fidelity conveyor-mode spin shuttling is considered the key enabler for scalable architectures of semiconductor quantum computers. We present confirmation of predicted spin-dephasing mechanisms, predict an alternative regime of high-fidelity shuttling, and support current strategies for advancing the fidelity of spin-qubit conveyor-mode shuttling.

\section*{Methods}

\subsection*{Device fabrication and design}
The device consists of a $\mathrm{Si}_{0.7}\mathrm{Ge}_{0.3}$/$^{28}\mathrm{Si}$/$\mathrm{Si}_{0.7}\mathrm{Ge}_{0.3}$ heterostructure grown on a silicon (001) wafer by CVD. It forms an electron quantum well in the \SI{5}{\nano \meter} thick, isotopically purified $^{28}$Si layer with <800\,ppm $^{29}$Si concentration (Fig.~\ref{fig:Setup_VS_map}b). The upper $\mathrm{Si}_{0.7}\mathrm{Ge}_{0.3}$ layer has a nominal thickness of  $30\,\mathrm{nm}$ and is passivated by an amorphous Si-rich cap~\cite{paqueletwuetz23}. The lower SiGe barrier is \SI{2.4}{\micro \meter} thick and grown on a relaxed, stepped SiGe buffer made of three equidistant \SI{1}{\micro \meter} thick steps. The tensile strained $^{28}$Si layer is significantly thinner than the Matthews-Blakeslee critical thickness for strain relaxation. This minimizes the chance of having relaxation-induced crystal defects at the critical quantum well interfaces and has the potential to increase valley splitting energies, as the wave function overlap with the SiGe barriers is increased~\cite{Wuetz2022}. Ohmic contacts are fabricated by phosphorus ion implantation at Helmholtz-Center Dresden-Rossendorf, followed by thermal activation at $730\,\mathrm{^{\circ}C}$ for $30\,\mathrm{s}$. The three metal gate layers fabricated at the Helmholtz Nano Facility \cite{Albrecht17} (called L1, L2 and L3 from bottom to top; positions in Fig.~\ref{fig:Setup_VS_map}b) are made of a $\mathrm{Ti}$/$\mathrm{Pt}$ ebeam-evaporated stack \cite{Volmer21} with thickness $5\,\mathrm{nm}$ and $15\,\mathrm{nm}$/$22\,\mathrm{nm}$/$29\,\mathrm{nm}$, respectively. They are shaped by 100\,keV electron-beam lithography and a metal lift-off process. They are electrically isolated from each other and the SiGe heterostructure by three $10\,\mathrm{nm}$ amorphous $\mathrm{Al}_2\mathrm{O}_3$ layers fabricated by atomic layer deposition.

The device design features two single electron transistors used as proximal charge sensors and electron reservoirs \cite{Klos18}  positioned at the ends of a one dimensional electron channel (1DEC) (Fig.~\ref{fig:Setup_VS_map}a). The 1DEC is formed by a \SI{1.2}{\micro \meter} split gate in the (110) crystallographic direction on the first metal layer (L1: SB, ST; purple) with \SI{200}{\nano \meter} spacing. It is covered by 17 narrow clavier gates distributed among metal layers L2 and L3 (L2: G2, S1, S3, G16; L3: G1, G3, S2, S4, G15, G17; dark magenta and bright red) with a combined gate pitch of \SI{70}{\nano \meter}. Importantly, all S-gates labeled the same are electrically connected outside the chip as sketched in Fig.~\ref{fig:Setup_VS_map}a, similar to Refs.~\cite{Seidler22,Xue23,Struck23,volmer24}. Thus, the clavier gates for the four gate sets S1-S4 are controlled by only four voltages. Each SET consists of three finger gates (L1: LP, RP; L2: LB1, LB2, RB1, RB2; yellow in Fig.~\ref{fig:Setup_VS_map}a) and one top gate (L3: LT, RT; orange). The shuttle operation is limited due to the rightmost gate of S4 being unintentionally disconnected.

\subsection*{Experimental setup}
The device is mounted on the mixing chamber stage of an Oxford Instruments dry dilution refrigerator at a temperature of $\approx60\,\mathrm{mK}$ and a global in-plane magnetic field generated by a superconducting magnet. All gates are connected via \SI{10}{\kilo \hertz} low-pass filtered loom lines (DC lines). Additionally, signals on G2, G3 as well as S1-S4 inside the refrigerator are carried to the sample PCB via coaxial cables with a bandwidth of \SI{20}{\giga \hertz} (RF lines) and are added to the DC lines with RC bias tees with a cutoff frequency of \SI{5}{\hertz}. At room temperature, DC voltages are generally applied using custom-made low-noise digital-to-analog converters. Pulsed voltages are applied by Zürich Instruments HDAWG arbitrary waveform generators on the RF lines. For each of the high frequency gates, the voltage pulses are also added to the DC lines in order to avoid transients from long measurements. The SET current is read-out by Basel Precision Instruments transimpedance amplifiers and AlazarTech digitizer cards. 

\subsection*{Efficiency of valley \add{splitting} mapping}
Mapping the local valley splitting is an efficient method for benchmarking heterostructures. Here, we estimate time requirements from state-of-the-art values for spin initialization and detection. The total measurement time $T_{\text{meas}}$ for a two-dimensional map is a product of the following parameters
\begin{equation}
\begin{aligned}
    T_{\text{meas}} =\tau_{ss}\cdot n_B\cdot & n_{\text{samples}} \cdot n_x \cdot n_y + \tau_{\text{B}},
\end{aligned}
\end{equation}

\noindent where $\tau_{ss}$ is the single-shot measurement time of PSB, $n_x=l_x/\delta_x$ and $n_y=l_y/\delta_y$ are the number of samples in parallel and perpendicular direction to the 1DEC, each given by the ratio of total size $l_i$ and resolution $\delta_i$. $n_B$ is the number of magnetic fields required to resolve the valley splittings clearly and $n_{\text{samples}}$ the number of repetitions to calculate $P_\text{S}$. $\tau_B$ is the time it takes to change to a different magnetic field.

\begin{table}[ht]
    \centering
    \begin{tabular}{cc}
        \textbf{Parameter} & \textbf{Value} \\
        \hline
        $\tau_{ss}$        & \SI{2}{\micro \second}~\cite{Stano22}  \\
        $n_B$              & 600 \\
        $n_{\text{samples}}$      & 100 \\
        $\tau_\text{B}$ & $\SI{1}{\second} \cdot n_B$ \\
        \hline
        $l_x$              &  \SI{10}{\micro \meter}~\cite{Xue23}  \\
        $\delta_x$         &  \SI{5}{\nano \meter} \\
        $n_x$              &  2000\\
        \hline
        $l_y$              &  \SI{40}{\nano \meter} \\
        $\delta_y$         &  \SI{5}{\nano \meter} \\
        $n_y$              &  8 \\
        \hline
        $T_{\text{meas}}$ &  \SI{2520}{\second}
    \end{tabular}
    \caption{State-of-art values for our time estimate of a two-dimensional valley splitting map of size $l_x \times l_y$.}
    \label{tab:time_requirements}
\end{table}

Our time estimate is based on values according to Tab.~\ref{tab:time_requirements}. We suggest a lateral spacing of \SI{5}{\nano \meter} as this is smaller than the quantum dot size. Hence, we arrive at 2000 sample points along the 1DEC and eight traces. The magnetic field needs to be swept at least in fields that cover the low valley splitting ranges below $\approx \SI{180}{\micro \electronvolt}$, so \SI{1800}{\milli \tesla} with \SI{3}{\milli \tesla} precision, thus $n_B=600$. We include \SI{1}{\second} for changing the magnetic field to the next value. Sufficient precision of $P_\text{S}$ is achieved by $n_\text{samples}=100$. Ideally, one would work with pulsed voltages on the long split gate defining the position of the 1DEC. This way, a retuning of the SET between displacements of the 1DEC (change of $y$) is not required. Since SET tuning is a matter of quality of automation, we omit it here. 

\add{We observe a slight drop in visibility of the signature of valley resonance, if the electron passes several spin-valley hotspots along its trajectory. Passing serveral spin-valley hotspots can be mitigated by shuttling faster and ultimately mitigated by optimizing the shuttle trajectory. We suggest the following algorithm to enlarge the area: First, an area of 40 nm by 500 nm is mapped and the best shuttle trajectory selected. This trajectory is used to reach the next area (of the same size as the first one) in order to perform mapping there.
Then a third area is added and the previous two are crossed by the optimal trajectory, and so on.}

In summary, a valley splitting map of a 10 micron shuttle device can be obtained fast to either find the best shuttle trajectory or to just benchmark heterostructures. The former can be further accelerated by a smart optimization algorithm that does not require measurement of the full 2D map. As the valley splitting map is encoded by the position of the Ge atoms in the quantum well region of a shuttle device, frequent recalibration is not necessary.   

\section*{Data availability}
The data that support the findings of this study are available in the \href{https://doi.org/10.5281/zenodo.17373163}{Zenodo repository}.

\section*{Acknowledgements}
We acknowledge valuable discussions with Merritt P. Losert and Mark Friesen and the support of the Dresden High Magnetic Field Laboratory (HLD) at the Helmholtz-Zentrum Dresden - Rossendorf (HZDR), member of the European Magnetic Field Laboratory (EMFL). The device fabrication has been done at HNF - Helmholtz Nano Facility, Research Center Juelich GmbH \cite{Albrecht17}.

\section*{Funding}
This work was funded by the German Research Foundation (DFG) within the project 289786932 (SCHR 1404/2-2) and under Germany's Excellence Strategy - Cluster of Excellence Matter and Light for Quantum Computing" (ML4Q) EXC 2004/2 - 390534769, and  by the European Union’s Horizon Research and Innovation Actions under Grant Agreement No. 101174557 (QLSI2). This research was sponsored in part
by The Netherlands Ministry of Defence under Awards
No. QuBits R23/009. The views, conclusions, and recommendations contained in this document are those of
the authors and are not necessarily endorsed nor should
they be interpreted as representing the official policies,
either expressed or implied, of The Netherlands Ministry
of Defence. The Netherlands Ministry of Defence is authorized to reproduce and distribute reprints for Government purposes notwithstanding any copyright notation
herein.

\section*{Author contributions}
M.V. set up and conducted the experiments assisted by T.S. Authors M.V., T.S.,  {\L}.C. and L.R.S. analyzed the data. The heterostructure was grown by D.D.E. and G.S. Device fabrication was done by J.T. and S.T and device pre-screening by A.S. Theory has been developed by {\L}.C with help of L.R.S and M.V. The experiment was designed and supervised by L.R.S. L.R.S. and H.B. provided guidance to all authors. M.V., {\L}.C. and L.R.S. wrote the manuscript which was commented on by all other authors.

\section*{Competing interests}
H.B, L.R.S., T.S. and M.V. are co-inventors of patent applications that cover conveyor-mode shuttling and/or its applications. G.S. is founding advisor of Groove Quantum BV and
declares equity interests. L.R.S. and H.B. are founders and shareholders of ARQUE Systems GmbH. The other authors declare no competing interest.


\begin{thebibliography}{10}
\expandafter\ifx\csname url\endcsname\relax
  \def\url#1{\texttt{#1}}\fi
\expandafter\ifx\csname urlprefix\endcsname\relax\def\urlprefix{URL }\fi
\providecommand{\bibinfo}[2]{#2}
\providecommand{\eprint}[2][]{\url{#2}}

\bibitem{Burkard23}
\bibinfo{author}{Burkard, G.}, \bibinfo{author}{Ladd, T.~D.}, \bibinfo{author}{Pan, A.}, \bibinfo{author}{Nichol, J.~M.} \& \bibinfo{author}{Petta, J.~R.}
\newblock \bibinfo{title}{Semiconductor spin qubits}.
\newblock \emph{\bibinfo{journal}{Rev. Mod. Phys.}} \textbf{\bibinfo{volume}{95}}, \bibinfo{pages}{025003} (\bibinfo{year}{2023}).

\bibitem{Struck2020}
\bibinfo{author}{Struck, T.} \emph{et~al.}
\newblock \bibinfo{title}{Low-frequency spin qubit energy splitting noise in highly purified $^{28}${Si/SiGe}}.
\newblock \emph{\bibinfo{journal}{npj Quantum Inf.}} \textbf{\bibinfo{volume}{6}}, \bibinfo{pages}{2056} (\bibinfo{year}{2020}).

\bibitem{philips22}
\bibinfo{author}{Philips, S. G.~J.} \emph{et~al.}
\newblock \bibinfo{title}{Universal control of a six-qubit quantum processor in silicon}.
\newblock \emph{\bibinfo{journal}{Nature}} \textbf{\bibinfo{volume}{609}}, \bibinfo{pages}{919--924} (\bibinfo{year}{2022}).

\bibitem{Stano22}
\bibinfo{author}{Stano, P.} \& \bibinfo{author}{Loss, D.}
\newblock \bibinfo{title}{Review of performance metrics of spin qubits in gated semiconducting nanostructures}.
\newblock \emph{\bibinfo{journal}{Nat. Rev. Phys.}} \textbf{\bibinfo{volume}{4}}, \bibinfo{pages}{672} (\bibinfo{year}{2022}).

\bibitem{defuentes25}
\bibinfo{author}{Fern\'andez~de Fuentes, I.} \emph{et~al.}
\newblock \bibinfo{title}{Running a six-qubit quantum circuit on a silicon spin-qubit array}.
\newblock \emph{\bibinfo{journal}{PRX Quantum}} \textbf{\bibinfo{volume}{7}}, \bibinfo{pages}{010308} (\bibinfo{year}{2026}).

\bibitem{Yoneda2018}
\bibinfo{author}{Yoneda, J.} \emph{et~al.}
\newblock \bibinfo{title}{A quantum-dot spin qubit with coherence limited by charge noise and fidelity higher than 99.9\%}.
\newblock \emph{\bibinfo{journal}{Nat. Nanotechnol.}} \textbf{\bibinfo{volume}{13}}, \bibinfo{pages}{102} (\bibinfo{year}{2018}).

\bibitem{Struck21}
\bibinfo{author}{Struck, T.} \emph{et~al.}
\newblock \bibinfo{title}{Robust and fast post-processing of single-shot spin qubit detection events with a neural network}.
\newblock \emph{\bibinfo{journal}{Sci. Rep.}} \textbf{\bibinfo{volume}{11}}, \bibinfo{pages}{16203} (\bibinfo{year}{2021}).

\bibitem{Noiri2022}
\bibinfo{author}{Noiri, A.} \emph{et~al.}
\newblock \bibinfo{title}{Fast universal quantum gate above the fault-tolerance threshold in silicon}.
\newblock \emph{\bibinfo{journal}{Nature}} \textbf{\bibinfo{volume}{601}}, \bibinfo{pages}{338} (\bibinfo{year}{2022}).

\bibitem{Mills2022}
\bibinfo{author}{Mills, A.~R.} \emph{et~al.}
\newblock \bibinfo{title}{Two-qubit silicon quantum processor with operation fidelity exceeding 99\%}.
\newblock \emph{\bibinfo{journal}{Sci. Adv.}} \textbf{\bibinfo{volume}{8}}, \bibinfo{pages}{eabn5130} (\bibinfo{year}{2022}).

\bibitem{Xue2022}
\bibinfo{author}{Xue, X.} \emph{et~al.}
\newblock \bibinfo{title}{Quantum logic with spin qubits crossing the surface code threshold}.
\newblock \emph{\bibinfo{journal}{Nature}} \textbf{\bibinfo{volume}{601}}, \bibinfo{pages}{343} (\bibinfo{year}{2022}).

\bibitem{Wu25}
\bibinfo{author}{Wu, Y.-H.} \emph{et~al.}
\newblock \bibinfo{title}{Simultaneous high-fidelity single-qubit gates in a spin qubit array}  (\bibinfo{year}{2025}).
\newblock \eprint{Arxiv preprint at https://arxiv.org/abs/2507.11918}.

\bibitem{Bluvstein22}
\bibinfo{author}{Bluvstein, D.} \emph{et~al.}
\newblock \bibinfo{title}{A quantum processor based on coherent transport of entangled atom arrays}.
\newblock \emph{\bibinfo{journal}{Nature}} \textbf{\bibinfo{volume}{604}}, \bibinfo{pages}{451--456} (\bibinfo{year}{2022}).

\bibitem{Arute19}
\bibinfo{author}{Arute, F.} \emph{et~al.}
\newblock \bibinfo{title}{Quantum supremacy using a programmable superconducting processor}.
\newblock \emph{\bibinfo{journal}{Nature}}  (\bibinfo{year}{2019}).

\bibitem{Monroe21}
\bibinfo{author}{Monroe, C.} \emph{et~al.}
\newblock \bibinfo{title}{Programmable quantum simulations of spin systems with trapped ions}.
\newblock \emph{\bibinfo{journal}{Rev. Mod. Phys.}} \textbf{\bibinfo{volume}{93}}, \bibinfo{pages}{025001} (\bibinfo{year}{2021}).

\bibitem{zwerver22}
\bibinfo{author}{Zwerver, A. M.~J.} \emph{et~al.}
\newblock \bibinfo{title}{Qubits made by advanced semiconductor manufacturing}.
\newblock \emph{\bibinfo{journal}{Nat. Electron.}} \textbf{\bibinfo{volume}{5}}, \bibinfo{pages}{184--190} (\bibinfo{year}{2022}).

\bibitem{neyens24}
\bibinfo{author}{Neyens, S.} \emph{et~al.}
\newblock \bibinfo{title}{Probing single electrons across 300-mm spin qubit wafers}.
\newblock \emph{\bibinfo{journal}{Nature}} \textbf{\bibinfo{volume}{629}}, \bibinfo{pages}{80--85} (\bibinfo{year}{2024}).

\bibitem{George25}
\bibinfo{author}{George, H.~C.} \emph{et~al.}
\newblock \bibinfo{title}{12-{Spin}-{Qubit} {Arrays} {Fabricated} on a 300 mm {Semiconductor} {Manufacturing} {Line}}.
\newblock \emph{\bibinfo{journal}{Nano Letters}} \textbf{\bibinfo{volume}{25}}, \bibinfo{pages}{793--799} (\bibinfo{year}{2025}).

\bibitem{Huckemann25}
\bibinfo{author}{Huckemann, T.} \emph{et~al.}
\newblock \bibinfo{title}{Industrially fabricated single-electron quantum dots in {Si/SiGe} heterostructures}.
\newblock \emph{\bibinfo{journal}{IEEE Electron Device Lett.}} \textbf{\bibinfo{volume}{46}}, \bibinfo{pages}{868--871} (\bibinfo{year}{2025}).

\bibitem{Vandersypen17}
\bibinfo{author}{Vandersypen, L. M.~K.} \emph{et~al.}
\newblock \bibinfo{title}{Interfacing spin qubits in quantum dots and donors---hot, dense, and coherent}.
\newblock \emph{\bibinfo{journal}{npj Quantum Inf.}} \textbf{\bibinfo{volume}{3}}, \bibinfo{pages}{34} (\bibinfo{year}{2017}).

\bibitem{Fujita17}
\bibinfo{author}{Fujita, T.}, \bibinfo{author}{Baart, T.~A.}, \bibinfo{author}{Reichl, C.}, \bibinfo{author}{Wegscheider, W.} \& \bibinfo{author}{Vandersypen, L. M.~K.}
\newblock \bibinfo{title}{Coherent shuttle of electron-spin states}.
\newblock \emph{\bibinfo{journal}{npj Quantum Inf.}} \textbf{\bibinfo{volume}{3}}, \bibinfo{pages}{22} (\bibinfo{year}{2017}).

\bibitem{Mills19}
\bibinfo{author}{Mills, A.~R.} \emph{et~al.}
\newblock \bibinfo{title}{Shuttling a single charge across a one-dimensional array of silicon quantum dots}.
\newblock \emph{\bibinfo{journal}{Nat. Commun.}} \textbf{\bibinfo{volume}{10}}, \bibinfo{pages}{1063} (\bibinfo{year}{2019}).

\bibitem{Yoneda21}
\bibinfo{author}{Yoneda, J.} \emph{et~al.}
\newblock \bibinfo{title}{Coherent spin qubit transport in silicon}.
\newblock \emph{\bibinfo{journal}{Nat. Commun.}} \textbf{\bibinfo{volume}{12}}, \bibinfo{pages}{4114} (\bibinfo{year}{2021}).

\bibitem{noiri22_2}
\bibinfo{author}{Noiri, A.} \emph{et~al.}
\newblock \bibinfo{title}{A shuttling-based two-qubit logic gate for linking distant silicon quantum processors}.
\newblock \emph{\bibinfo{journal}{Nat. Commun.}} \textbf{\bibinfo{volume}{13}}, \bibinfo{pages}{5740} (\bibinfo{year}{2022}).

\bibitem{Zwerver23}
\bibinfo{author}{Zwerver, A.} \emph{et~al.}
\newblock \bibinfo{title}{Shuttling an electron spin through a silicon quantum dot array}.
\newblock \emph{\bibinfo{journal}{PRX Quantum}} \textbf{\bibinfo{volume}{4}}, \bibinfo{pages}{030303} (\bibinfo{year}{2023}).

\bibitem{foster24}
\bibinfo{author}{Foster, N.~D.}, \bibinfo{author}{Henshaw, J.~D.}, \bibinfo{author}{Rudolph, M.}, \bibinfo{author}{Luhman, D.~R.} \& \bibinfo{author}{Jock, R.~M.}
\newblock \bibinfo{title}{Dephasing and error dynamics affecting a singlet–triplet qubit during coherent spin shuttling}.
\newblock \emph{\bibinfo{journal}{npj Quantum Inf.}} \textbf{\bibinfo{volume}{11}}, \bibinfo{pages}{63} (\bibinfo{year}{2025}).

\bibitem{Taylor05}
\bibinfo{author}{Taylor, J.~M.} \emph{et~al.}
\newblock \bibinfo{title}{Fault-tolerant architecture for quantum computation using electrically controlled semiconductor spins}.
\newblock \emph{\bibinfo{journal}{Nat. Phys.}} \textbf{\bibinfo{volume}{1}}, \bibinfo{pages}{177--183} (\bibinfo{year}{2005}).

\bibitem{Boter22}
\bibinfo{author}{Boter, J.~M.} \emph{et~al.}
\newblock \bibinfo{title}{Spiderweb array: A sparse spin-qubit array}.
\newblock \emph{\bibinfo{journal}{Phys. Rev. Appl.}} \textbf{\bibinfo{volume}{18}}, \bibinfo{pages}{024053} (\bibinfo{year}{2022}).

\bibitem{Kuenne23}
\bibinfo{author}{Künne, M.} \emph{et~al.}
\newblock \bibinfo{title}{The {SpinBus} architecture for scaling spin qubits with electron shuttling}.
\newblock \emph{\bibinfo{journal}{Nat. Commun.}} \textbf{\bibinfo{volume}{15}}, \bibinfo{pages}{4977} (\bibinfo{year}{2024}).

\bibitem{Ginzel24}
\bibinfo{author}{Ginzel, F.} \emph{et~al.}
\newblock \bibinfo{title}{Scalable parity architecture with a shuttling-based spin qubit processor}.
\newblock \emph{\bibinfo{journal}{Phys. Rev. B}} \textbf{\bibinfo{volume}{110}}, \bibinfo{pages}{075302} (\bibinfo{year}{2024}).

\bibitem{Yenilen25}
\bibinfo{author}{Yenilen, B.}, \bibinfo{author}{Sala, A.}, \bibinfo{author}{Bluhm, H.}, \bibinfo{author}{Müller, M.} \& \bibinfo{author}{Rispler, M.}
\newblock \bibinfo{title}{Performance of the spin qubit shuttling architecture for a surface code implementation} (\bibinfo{year}{2025}).
\newblock \eprint{Preprint at https://arxiv.org/abs/2503.10601}.

\bibitem{Xue23}
\bibinfo{author}{Xue, R.} \emph{et~al.}
\newblock \bibinfo{title}{Si/{SiGe} {QuBus} for single electron information-processing devices with memory and micron-scale connectivity function}.
\newblock \emph{\bibinfo{journal}{Nat. Commun.}} \textbf{\bibinfo{volume}{15}}, \bibinfo{pages}{2296} (\bibinfo{year}{2024}).

\bibitem{Beer25}
\bibinfo{author}{Beer, M.} \emph{et~al.}
\newblock \bibinfo{title}{Conveyor-mode electron shuttling through a {T}-junction in {Si/SiGe}} (\bibinfo{year}{2026}).
\newblock \eprint{Preprint at https://arxiv.org/abs/2601.03942}.

\bibitem{Seidler22}
\bibinfo{author}{Seidler, I.} \emph{et~al.}
\newblock \bibinfo{title}{{Conveyor-mode single-electron shuttling in {Si/SiGe} for a scalable quantum computing architecture}}.
\newblock \emph{\bibinfo{journal}{npj Quantum Inf.}} \textbf{\bibinfo{volume}{8}}, \bibinfo{pages}{100} (\bibinfo{year}{2022}).

\bibitem{Langrock23}
\bibinfo{author}{Langrock, V.} \emph{et~al.}
\newblock \bibinfo{title}{Blueprint of a scalable spin qubit shuttle device for coherent mid-range qubit transfer in disordered {Si/SiGe/SiO}$_{2}$}.
\newblock \emph{\bibinfo{journal}{PRX Quantum}} \textbf{\bibinfo{volume}{4}}, \bibinfo{pages}{020305} (\bibinfo{year}{2023}).

\bibitem{Zhao25}
\bibinfo{author}{Zhao, Q.-T.} \emph{et~al.}
\newblock \bibinfo{title}{Ultra-low-power cryogenic complementary metal oxide semiconductor technology}.
\newblock \emph{\bibinfo{journal}{Nat. Rev. Electr. Eng.}} \textbf{\bibinfo{volume}{2}}, \bibinfo{pages}{277--290} (\bibinfo{year}{2025}).

\bibitem{Struck23}
\bibinfo{author}{Struck, T.} \emph{et~al.}
\newblock \bibinfo{title}{Spin-{EPR}-pair separation by conveyor-mode single electron shuttling in {Si}/{SiGe}}.
\newblock \emph{\bibinfo{journal}{Nat. Commun.}} \textbf{\bibinfo{volume}{15}}, \bibinfo{pages}{1325} (\bibinfo{year}{2024}).

\bibitem{desmet24}
\bibinfo{author}{De~Smet, M.} \emph{et~al.}
\newblock \bibinfo{title}{High-fidelity single-spin shuttling in silicon}.
\newblock \emph{\bibinfo{journal}{Nat. Nanotechnol.}} \textbf{\bibinfo{volume}{20}}, \bibinfo{pages}{866--872} (\bibinfo{year}{2025}).


\bibitem{matsumoto25}
\bibinfo{author}{Matsumoto~Y.} \emph{et~al.}   
\newblock \bibinfo{title}{Two-qubit logic and teleportation with mobile spin qubits in silicon}. 
\newblock \emph{\bibinfo{journal}{Nature}} \textbf{\bibinfo{volume}{653}}, \bibinfo{pages}{391--397} (\bibinfo{year}{2026}).

\bibitem{Lima24}
\bibinfo{author}{Lima, J. R.~F.} \& \bibinfo{author}{Burkard, G.}
\newblock \bibinfo{title}{Valley splitting depending on the size and location of a silicon quantum dot}.
\newblock \emph{\bibinfo{journal}{Phys. Rev. Mater.}} \textbf{\bibinfo{volume}{8}}, \bibinfo{pages}{036202} (\bibinfo{year}{2024}).

\bibitem{Lima25}
\bibinfo{author}{Lima, J. R.~F.} \& \bibinfo{author}{Burkard, G.}
\newblock \bibinfo{title}{Partial landau-zener transitions and applications to qubit shuttling}.
\newblock \emph{\bibinfo{journal}{Phys. Rev. B}} \textbf{\bibinfo{volume}{111}}, \bibinfo{pages}{235439} (\bibinfo{year}{2025}).

\bibitem{Hollmann20}
\bibinfo{author}{Hollmann, A.} \emph{et~al.}
\newblock \bibinfo{title}{Large, tunable valley splitting and single-spin relaxation mechanisms in a {Si}/{Si}$_{x}${Ge}$_{1\ensuremath{-}x}$ quantum dot}.
\newblock \emph{\bibinfo{journal}{Phys. Rev. Appl.}} \textbf{\bibinfo{volume}{13}}, \bibinfo{pages}{034068} (\bibinfo{year}{2020}).

\bibitem{Wuetz2022}
\bibinfo{author}{Paquelet~Wuetz, B.} \emph{et~al.}
\newblock \bibinfo{title}{Atomic fluctuations lifting the energy degeneracy in {Si}/{SiGe} quantum dots}.
\newblock \emph{\bibinfo{journal}{Nat. Commun.}} \textbf{\bibinfo{volume}{13}}, \bibinfo{pages}{7730} (\bibinfo{year}{2022}).

\bibitem{volmer24}
\bibinfo{author}{Volmer, M.} \emph{et~al.}
\newblock \bibinfo{title}{Mapping of valley splitting by conveyor-mode spin-coherent electron shuttling}.
\newblock \emph{\bibinfo{journal}{npj Quantum Inf.}} \textbf{\bibinfo{volume}{10}}, \bibinfo{pages}{61} (\bibinfo{year}{2024}).

\bibitem{Marcks25}
\bibinfo{author}{Marcks, J.~C.} \emph{et~al.}
\newblock \bibinfo{title}{Valley splitting correlations across a silicon quantum well containing germanium}.
\newblock \emph{\bibinfo{journal}{Nat. Commun.}} \textbf{\bibinfo{volume}{16}}, \bibinfo{pages}{11381} (\bibinfo{year}{2025}).

\bibitem{Klos24}
\bibinfo{author}{Klos, J.} \emph{et~al.}
\newblock \bibinfo{title}{Atomistic compositional details and their importance for spin qubits in isotope-purified silicon quantum wells}.
\newblock \emph{\bibinfo{journal}{Adv. Sci.}} \textbf{\bibinfo{volume}{11}}, \bibinfo{pages}{2407442} (\bibinfo{year}{2024}).

\bibitem{thayil25}
\bibinfo{author}{Thayil, A.}, \bibinfo{author}{Ermoneit, L.} \& \bibinfo{author}{Kantner, M.}
\newblock \bibinfo{title}{Theory of valley splitting in {Si/SiGe} spin qubits: Interplay of strain, resonances, and random alloy disorder}.
\newblock \emph{\bibinfo{journal}{Phys. Rev. B}} \textbf{\bibinfo{volume}{112}}, \bibinfo{pages}{115303} (\bibinfo{year}{2025}).

\bibitem{Friesen2007}
\bibinfo{author}{Friesen, M.}, \bibinfo{author}{Chutia, S.}, \bibinfo{author}{Tahan, C.} \& \bibinfo{author}{Coppersmith, S.~N.}
\newblock \bibinfo{title}{Valley splitting theory of $\mathrm{Si}\mathrm{Ge}/\mathrm{Si}/\mathrm{Si}\mathrm{Ge}$ quantum wells}.
\newblock \emph{\bibinfo{journal}{Phys. Rev. B}} \textbf{\bibinfo{volume}{75}}, \bibinfo{pages}{115318} (\bibinfo{year}{2007}).

\bibitem{Losert23}
\bibinfo{author}{Losert, M.~P.} \emph{et~al.}
\newblock \bibinfo{title}{Practical strategies for enhancing the valley splitting in {Si/SiGe} quantum wells}.
\newblock \emph{\bibinfo{journal}{Phys. Rev. B}} \textbf{\bibinfo{volume}{108}}, \bibinfo{pages}{125405} (\bibinfo{year}{2023}).

\bibitem{woods24}
\bibinfo{author}{Woods, B.~D.} \emph{et~al.}
\newblock \bibinfo{title}{Coupling conduction-band valleys in {SiGe} heterostructures via shear strain and {Ge} concentration oscillations}.
\newblock \emph{\bibinfo{journal}{npj Quantum Inf.}} \textbf{\bibinfo{volume}{10}}, \bibinfo{pages}{54} (\bibinfo{year}{2024}).

\bibitem{Kawakami2014}
\bibinfo{author}{Kawakami, E.} \emph{et~al.}
\newblock \bibinfo{title}{Electrical control of a long-lived spin qubit in a {Si/SiGe} quantum dot}.
\newblock \emph{\bibinfo{journal}{Nat. Nanotechnol.}} \textbf{\bibinfo{volume}{9}}, \bibinfo{pages}{666} (\bibinfo{year}{2014}).

\bibitem{Losert24}
\bibinfo{author}{Losert, M.~P.} \emph{et~al.}
\newblock \bibinfo{title}{Strategies for enhancing spin-shuttling fidelities in $\mathrm{Si}$/$\mathrm{Si}$$\mathrm{Ge}$ quantum wells with random-alloy disorder}.
\newblock \emph{\bibinfo{journal}{PRX Quantum}} \textbf{\bibinfo{volume}{5}}, \bibinfo{pages}{040322} (\bibinfo{year}{2024}).

\bibitem{David24}
\bibinfo{author}{David, A.} \emph{et~al.}
\newblock \bibinfo{title}{Long distance spin shuttling enabled by few-parameter velocity optimization}  (\bibinfo{year}{2024}).
\newblock \eprint{Preprint at https://arxiv.org/abs/2409.07600}.

\bibitem{Nemeth24}
\bibinfo{author}{N\'emeth, R.} \emph{et~al.}
\newblock \bibinfo{title}{Omnidirectional shuttling to avoid valley excitations in $\mathrm{Si}$/$\mathrm{Si}\mathrm{Ge}$ quantum wells}.
\newblock \emph{\bibinfo{journal}{PRX Quantum}} \textbf{\bibinfo{volume}{7}}, \bibinfo{pages}{010336} (\bibinfo{year}{2026}).

\bibitem{paqueletwuetz23}
\bibinfo{author}{Paquelet~Wuetz, B.} \emph{et~al.}
\newblock \bibinfo{title}{Reducing charge noise in quantum dots by using thin silicon quantum wells}.
\newblock \emph{\bibinfo{journal}{Nat. Commun.}} \textbf{\bibinfo{volume}{14}}, \bibinfo{pages}{1385} (\bibinfo{year}{2023}).

\bibitem{Cywinski25}
\bibinfo{author}{Cywinski, {\L}.}
\emph{et~al.}
\newblock \bibinfo{title}{Singlet-triplet oscillations in multivalley {Si} quantum dots}  (\bibinfo{year}{2026}).
\newblock \eprint{In preparation}.

\bibitem{Chen21}
\bibinfo{author}{Chen, E.~H.} \emph{et~al.}
\newblock \bibinfo{title}{Detuning axis pulsed spectroscopy of valley-orbital states in {Si}/{Si}-{Ge} quantum dots}.
\newblock \emph{\bibinfo{journal}{Phys. Rev. Appl.}} \textbf{\bibinfo{volume}{15}}, \bibinfo{pages}{044033} (\bibinfo{year}{2021}).

\bibitem{Dyakonov72}
\bibinfo{author}{D'yakonov, M.~I.} \& \bibinfo{author}{Perel', V.~I.}
\newblock \bibinfo{title}{Spin relaxation of conduction electrons in noncentrosymmetric semiconductors}.
\newblock \emph{\bibinfo{journal}{Soviet Physics Solid State}} \textbf{\bibinfo{volume}{13}}, \bibinfo{pages}{3023--3026} (\bibinfo{year}{1972}).
\newblock \bibinfo{note}{Translated from Fiz. Tverd. Tela 13, 3581--3585 (1971)}.

\bibitem{Zutic04}
\bibinfo{author}{\ifmmode \check{Z}\else \v{Z}\fi{}uti\ifmmode~\acute{c}\else \'{c}\fi{}, I.}, \bibinfo{author}{Fabian, J.} \& \bibinfo{author}{Das~Sarma, S.}
\newblock \bibinfo{title}{Spintronics: Fundamentals and applications}.
\newblock \emph{\bibinfo{journal}{Rev. Mod. Phys.}} \textbf{\bibinfo{volume}{76}}, \bibinfo{pages}{323--410} (\bibinfo{year}{2004}).

\bibitem{volmer26}
\bibinfo{author}{Volmer, M.} \emph{et~al.}
\newblock \bibinfo{title}{Mapping g-factors and complex intervalley coupling in {Si/SiGe} by conveyor-mode shuttling} (\bibinfo{year}{2026}).
\newblock \eprint{Preprint at https://arxiv.org/abs/2603.01844}.

\bibitem{Lima23}
\bibinfo{author}{Lima, J. R.~F.} \& \bibinfo{author}{Burkard, G.}
\newblock \bibinfo{title}{Interface and electromagnetic effects in the valley splitting of {Si} quantum dots}.
\newblock \emph{\bibinfo{journal}{Mater. Quantum Technol.}} \textbf{\bibinfo{volume}{3}}, \bibinfo{pages}{025004} (\bibinfo{year}{2023}).

\bibitem{Oda24}
\bibinfo{author}{Oda, Y.}, \bibinfo{author}{Losert, M.~P.} \& \bibinfo{author}{Kestner, J.~P.}
\newblock \bibinfo{title}{Suppressing si valley excitation and valley-induced spin dephasing for long-distance shuttling}.
\newblock \emph{\bibinfo{journal}{Phys. Rev. Lett.}} \textbf{\bibinfo{volume}{136}}, \bibinfo{pages}{020802} (\bibinfo{year}{2026}).

\bibitem{Pazhedath24}
\bibinfo{author}{Pazhedath, A.~M.} \emph{et~al.}
\newblock \bibinfo{title}{Large spin-shuttling oscillations enabling high-fidelity single-qubit gates}.
\newblock \emph{\bibinfo{journal}{Phys. Rev. Appl.}} \textbf{\bibinfo{volume}{24}}, \bibinfo{pages}{034029} (\bibinfo{year}{2025}).

\bibitem{McJunkin21}
\bibinfo{author}{McJunkin, T.} \emph{et~al.}
\newblock \bibinfo{title}{Valley splittings in {Si}/{SiGe} quantum dots with a germanium spike in the silicon well}.
\newblock \emph{\bibinfo{journal}{Phys. Rev. B}} \textbf{\bibinfo{volume}{104}}, \bibinfo{pages}{085406} (\bibinfo{year}{2021}).

\bibitem{McJunkin22}
\bibinfo{author}{McJunkin, T.} \emph{et~al.}
\newblock \bibinfo{title}{{SiGe} quantum wells with oscillating {Ge} concentrations for quantum dot qubits}.
\newblock \emph{\bibinfo{journal}{Nat. Commun.}} \textbf{\bibinfo{volume}{13}}, \bibinfo{pages}{7777} (\bibinfo{year}{2022}).

\bibitem{Gradwohl25}
\bibinfo{author}{Gradwohl, K.-P.} \emph{et~al.}
\newblock \bibinfo{title}{Enhanced nanoscale {Ge} concentration oscillations in {Si}/{SiGe} quantum well through controlled segregation}.
\newblock \emph{\bibinfo{journal}{Nano Lett.}} \textbf{\bibinfo{volume}{25}}, \bibinfo{pages}{4204--4210} (\bibinfo{year}{2025}).

\bibitem{Albrecht17}
\bibinfo{author}{Albrecht, W.}, \bibinfo{author}{Moers, J.} \& \bibinfo{author}{Hermanns, B.}
\newblock \bibinfo{title}{{HNF} - {Helmholtz} {Nano} {Facility}}.
\newblock \emph{\bibinfo{journal}{J. Large Scale Res. Facil. \mbox{(JLSRF)}}} \textbf{\bibinfo{volume}{3}}, \bibinfo{pages}{A112} (\bibinfo{year}{2017}).

\bibitem{Volmer21}
\bibinfo{author}{Volmer, F.} \emph{et~al.}
\newblock \bibinfo{title}{How to solve problems in micro- and nanofabrication caused by the emission of electrons and charged metal atoms during e-beam evaporation}.
\newblock \emph{\bibinfo{journal}{Journal of Physics D: Applied Physics}} \textbf{\bibinfo{volume}{54}}, \bibinfo{pages}{225304} (\bibinfo{year}{2021}).

\bibitem{Klos18}
\bibinfo{author}{Klos, J.}, \bibinfo{author}{Hassler, F.}, \bibinfo{author}{Cerfontaine, P.}, \bibinfo{author}{Bluhm, H.} \& \bibinfo{author}{Schreiber, L.~R.}
\newblock \bibinfo{title}{Calculation of tunnel couplings in open gate-defined disordered quantum dot systems}.
\newblock \emph{\bibinfo{journal}{Phys. Rev. B}} \textbf{\bibinfo{volume}{98}}, \bibinfo{pages}{155320} (\bibinfo{year}{2018}).

\end{thebibliography}
\end{document}